\documentstyle[preprint,aps]{revtex}
\begin{document}

\draft
\title
{Fractal distribution function and fractal-deformed 
Heisenberg algebras}

\author
{Wellington da Cruz\footnote{E-mail: wdacruz@exatas.uel.br}}

\address
{Departamento de F\'{\i}sica,\\
 Universidade Estadual de Londrina, Caixa Postal 6001,\\
Cep 86051-970 Londrina, PR, Brazil\\}
 
\date{\today}

\maketitle

\begin{abstract}

We consider the concept of {\it fractons}, i.e. particles or 
quasiparticles which obey specific {\it fractal distribution function} 
and for each 
{\it universal class $h$ of particles} we obtain a fractal-deformed Heisenberg 
algebra. This one takes into account the braid group structure 
of these objects which live in two-dimensional multiply connected space.
 
\end{abstract}

%\narrowtext

The concept of {\it fractons} as objects which live in two-dimensional 
multiply connected space, satisfying {\it fractal distribution function } 
was introduced in\cite{R1}. There and another places, we have exploited 
these ideas in the contexts of Fractional quantum Hall effect, 
High-$T_{c}$ superconductivity\cite{R2}, 
Conformal field theories-CFT\cite{R3}, Luttinger liquids\cite{R4} and 
Black hole entropy\cite{R5}. A beautiful connection between {\bf Number Theory 
and Physics} came to out through our discussion of these topics: Farey 
series of rational numbers, Rogers dilogarithm function, fractal functions, 
algebraic equations are related, in some sense,  
to the notion of Hausdorff 
dimension\cite{R1,R3}.

We have defined {\it universal classes $h$ of particles or quasiparticles} which 
carry rational or irrational spin quantum number $s$. Thus for the 
fractal parameter or Hausdorff dimension $h$ defined into the 
interval $1$$\;$$ < $$\;$$h$$\;$$ <$$\;$$ 2$, each class gets its specific 
distribution function from {\it a unique} expression, the {\bf Fractal distribution 
function}\cite{R1}

\begin{eqnarray}
\label{e.h} 
n=\frac{1}{{\cal{Y}}[\xi]-h}
\end{eqnarray}

\noindent where the function ${\cal{Y}}[\xi]$ satisfies 
the equation 

\begin{eqnarray}
\xi=\biggl\{{\cal{Y}}[\xi]-1\biggr\}^{h-1}
\biggl\{{\cal{Y}}[\xi]-2\biggr\}^{2-h}.
\end{eqnarray}

\noindent We also have 
 
\begin{eqnarray}
\xi^{-1}=\biggl\{\Theta[{\cal{Y}}]\biggr\}^{h-2}-
\biggl\{\Theta[{\cal{Y}}]\biggr\}^{h-1}
\end{eqnarray}

\noindent where $\xi=\exp\left\{(\epsilon-\mu)/KT\right\}$ has the usual 
definition and 

\begin{eqnarray}
\Theta[{\cal{Y}}]=
\frac{{\cal{Y}}[\xi]-2}{{\cal{Y}}[\xi]-1}
\end{eqnarray}

\noindent is the {\it single-particle 
partition function}. In this way {\it the fermionic} ($h=1$) {\it and 
bosonic} ($h=2$) {\it distributions are generalized naturally}. 
On the other hand, the free energy for particles in a given 
quantum state is expressed as 

\begin{eqnarray}
{\cal{F}}[h]=KT\ln\Theta[{\cal{Y}}]
\end{eqnarray}
 
\noindent  such that for fermions 

\begin{eqnarray}
{\cal{F}}[1]=-KT\ln\left\{1+\xi^{-1}\right\}
\end{eqnarray}

\noindent and for bosons 

\begin{eqnarray}
{\cal{F}}[2]=KT\ln\left\{1-\xi^{-1}\right\}.
\end{eqnarray}

\noindent Hence we obtain the mean occupation number

\begin{eqnarray}
\label{e.h} 
n[h]&=&\xi\frac{\partial}{\partial{\xi}}\ln\Theta[{\cal{Y}}]
=\frac{1}{KT}\xi\frac{\partial\;{\cal{F}}}{\partial{\xi}}.
\end{eqnarray}

\noindent Another aspect of our formulation is that the classes 
$h$ satisfy 
a {\it duality simmetry} defined by ${\tilde{h}}=3-h$, such 
that bosons and fermions are dual objects, 
and as a consequence we extract a 
{\it fractal supersymmetry} which defines pairs of particles $\left(s,s+
\frac{1}{2}\right)$. This way, {\it the fractal distribution function 
is understood as a} 
{\bf quantum-geometrical} {\it description of the statistical laws of Nature, 
since it encodes the fractal characteristic of the quantum path, which 
reflects the Heisenberg uncertainty principle} .

A {\it fractal index} is asssociated with each class and defined by\cite{R3}

\begin{equation}
\label{e.1}
i_{f}[h]=\frac{6}{\pi^2}\int_{\infty(T=0)}^{1(T=\infty)}
\frac{d\xi}{\xi}\ln\left\{\Theta[\cal{Y}(\xi)]\right\}
\end{equation}

\noindent so we obtain for the bosonic class $i_{f}[2]=1$, for the 
fermionic class $i_{f}[1]=\frac{1}{2}$ and for the universal class 
$h=\frac{3}{2}$, we have $i_{f}[\frac{3}{2}]=\frac{3}{5}$. Thus for 
the interval of definition  $ 1$$\;$$ \leq $$\;$$h$$\;$$ \leq $$\;$$ 2$, 
there exists the correspondence $\frac{1}{2}$$\;$$ 
\leq $$\;$$i_{f}[h]$$\;$$ \leq $$\;$$ 1$, which signalizes the connection 
between fractons and CFT-quasiparticles ( edge excitations ), 
in accordance 
with the unitary
$c$$\;$$ <$$\;$$ 1$ representations of the central 
charge ( a dimensionless number which characterizes that 
two-dimensional field theories )\cite{R6}.

The particles of each class are collected taking 
into account the {\it fractal spectrum}

\begin{eqnarray}
&&h-1=1-\nu,\;\;\;\; 0 < \nu < 1;\;\;\;\;\;\;\;\;
 h-1=\nu-1,\;
\;\;\;\;\;\; 1 <\nu < 2;\\
&&etc.\nonumber
\end{eqnarray}

\noindent and the spin-statistics relation $\nu=2s$. For example, 
consider the universal classes with distinct values of spin $
\biggl\{\frac{1}{2},\frac{3}{2},\frac{5}{2},\cdots\biggr\}_{h=1}$, $
\biggl\{0,1,2,\cdots\biggr\}_{h=2}$ and $
\biggl\{\frac{1}{4},\frac{3}{4},\frac{5}{4},\cdots\biggr\}_{h=\frac{3}{2}}$, 
then we have the Fermi-Dirac distribution

\begin{eqnarray}
n[1]=\frac{1}{\xi+1}
\end{eqnarray}

\noindent the Bose-Einstein distribution

\begin{eqnarray}
n[2]=\frac{1}{\xi-1}
\end{eqnarray}

\noindent and the Fractal  distribution

\begin{eqnarray}
n\left[\frac{3}{2}\right]=\frac{1}{\sqrt{\frac{1}{4}+\xi^2}}.
\end{eqnarray}

\noindent Now, we introduce the notion of a {\it fractal-deformed Heisenberg algebra} 
for each universal class $h$ of particles, which generalizes 
the fermionic and bosonic ones. As we noted before, fractons are 
objects which live in two-dimensional multiply connected space and so, 
we have a braid group structure behind the process of exchange of any two 
of these.

The fractal-deformed Heisenberg 
algebra is obtained of the relation

\begin{eqnarray}
{\bf a}(x){\bf a}^{\dagger}(y)-f[\pm h]{\bf a}^{\dagger}(y)
{\bf a}(x)=\delta (x-y),
\end{eqnarray}

\noindent between creation and annihilation operators. The factor of 
deformation is defined as 

\begin{eqnarray}
f[\pm h]=exp\left(\pm\imath h\pi\right), 
\end{eqnarray}

\noindent such that for $h=1$ and $x=y$, we reobtain the fermionic 
anticommutation relations $\left\{{\bf a}(x),
{\bf a}^{\dagger}(x)\right\}=1$, and for  $h=2$ and $x=y$, we 
reobtain the bosonic 
commutation relations $\left[{\bf a}(x),
{\bf a}^{\dagger}(x)\right]=1$. If $x\neq y$ and 
$1$$\;$$ < $$\;$$h$$\;$$ <$$\;$$ 2$, we have nonlocal operators 
for fractons

\begin{eqnarray}
{\bf a}(x){\bf a}^{\dagger}(y)=f[\pm h]{\bf a}^{\dagger}(y){\bf a}(x).
\end{eqnarray}

\noindent The braiding relations have the hermiticities

\begin{eqnarray}
{\bf a}(x){\bf a}^{\dagger}(y)&=&f[+h]{\bf a}^{\dagger}(y){\bf a}(x)\\
{\bf a}^{\dagger}(x){\bf a}(y)&=&f[+h]{\bf a}(y){\bf a}^{\dagger}(x)\\
\nonumber\\
{\bf a}(x){\bf a}^{\dagger}(y)&=&f[-h]{\bf a}^{\dagger}(y){\bf a}(x)\\
{\bf a}^{\dagger}(x){\bf a}(y)&=&f[-h]{\bf a}(y){\bf a}^{\dagger}(x)\\
\nonumber\\
{\bf a}(x){\bf a}(y)&=&f[+h]{\bf a}(y){\bf a}(x)\\
{\bf a}^{\dagger}(x){\bf a}^{\dagger}(y)&=&f[+h]{\bf a}^{\dagger}(y)
{\bf a}^{\dagger}(x)\\
\nonumber\\
{\bf a}(x){\bf a}(y)&=&f[-h]{\bf a}(y){\bf a}(x)\\
{\bf a}^{\dagger}(x){\bf a}^{\dagger}(y)&=&f[-h]{\bf a}^{\dagger}(y)
{\bf a}^{\dagger}(x),
\end{eqnarray}

\noindent where $f[+h]$ stands for anticlockwise exchange and $f[-h]$ for 
clockwise exchange.

\noindent The statistics parameter $\nu$ for the plus sign has the pattern 

\begin{eqnarray}
\left\{-,+,-,+,\cdots\right\}_{h}
\end{eqnarray}

\noindent and the minus sign, gives us another one

\begin{eqnarray}
\left\{+,-,+,-,\cdots\right\}_{h}.
\end{eqnarray}

\noindent For example, consider the class $h=\frac{3}{2}$, then

\begin{eqnarray}
\biggl\{\mp\frac{1}{2},\pm\frac{3}{2},\mp\frac{5}{2},
\pm\frac{7}{2},\cdots\biggr\}_{h=\frac{3}{2}},
\end{eqnarray}

\noindent i.e. 

\begin{eqnarray}
f[+h]=e^{-\imath \frac{1}{2}\pi}=e^{+\imath \frac{3}{2}\pi}
=e^{-\imath \frac{5}{2}\pi}=e^{+\imath \frac{7}{2}\pi}=\cdots
\end{eqnarray}

\noindent and

\begin{eqnarray}
f[-h]=e^{+\imath \frac{1}{2}\pi}=e^{-\imath \frac{3}{2}\pi}
=e^{+\imath \frac{5}{2}\pi}=e^{-\imath \frac{7}{2}\pi}=\cdots.
\end{eqnarray}

\noindent As we can see, in each class, the particles with different 
values of statistics parameter ($\nu=2s$) 
have unambiguously distinct braiding properties to obey 
the fractal distribution function determined by the parameter $h$. 

The phase of the wave function ( consider two-particle system ) 
changes by $+\pi\nu$ ( anti-clockwise exchange ) and  $-\pi\nu$  
( clockwise exchange ) in response to which way we braid in interchanging 
$x$ and $y$. On the one hand, the violation of the discrete symmetries of parity 
and time reversal is verified for such fractons, i.e. 

\begin{eqnarray}
 e^{\pm\imath\pi\nu}\longrightarrow e^{\mp\imath\pi\nu}.
\end{eqnarray}

To summarize, we have determined the fractal-deformed 
Heisenberg algebras for universal classes $h$ of fractons 
which obey specific {\it fractal distribution function}\cite{R1}. 
This one {\it is just a natural } ( {\it besides its simplicity and elegance} ) 
{\it generalization of the fermionic and bosonic distributions 
for particles with braiding properties}. A next step is to consider 
the construction of the nonlocal operators for fractons 
taking into account an angle function as discussed, for example, 
by Lerda and Sciuto 
in\cite{R7}.

\end{document}